\documentclass[acmlarge,screen,nonacm]{acmart}

\usepackage{longtable}
\usepackage{natbib}

\usepackage{listings}
\usepackage{xcolor}
\lstdefinestyle{python}{
    language=Python,                 
    backgroundcolor=\color{white},   
    basicstyle=\ttfamily\small,      
    keywordstyle=\color{blue},        
    morekeywords={'def', '='}, 
    commentstyle=\color{green!70!black}, 
    stringstyle=\color{black},         
    showstringspaces=false,          
    numbers=left,                    
    numberstyle=\tiny\color{gray},   
    stepnumber=1,                    
    numbersep=5pt,                    
    escapeinside={(*@}{@*)},             
}
\usepackage{multirow}
\usepackage[utf8]{inputenc}
\usepackage{tcolorbox}
\tcbuselibrary{breakable}
\usepackage{lipsum} 
\usepackage{titlesec} 
\newtcolorbox{promptbox}[1][]{breakable, colback=white, colframe=gray!80, fonttitle=\bfseries, title=#1}

\usepackage{color}
\definecolor{YC}{RGB}{255, 99, 71} 

\begin{document}

\title{An LLM Agent for Automatic Geospatial Data Analysis}
\thanks{This work is supported by \textit{Agence Nationale de la Recherche} (ANR) under the ANR-21-CE23-0011 project. The GitHub repository for this work will be made available at: https://github.com/Yusin2Chen/GeoAgent}
\author{Yuxing Chen}
\email{chenyuxing16@mails.ucas.ac.cn}
\author{Weijie Wang}
\author{Sylvain Lobry}
\author{Camille Kurtz}





\renewcommand{\shortauthors}{Chen and Wang, et al.}

\begin{abstract}
  Large language models (LLMs) are being used in data science code generation tasks, but they often struggle with complex sequential tasks, leading to logical errors.
  Their application to geospatial data processing is particularly challenging due to difficulties in incorporating complex data structures and spatial constraints, effectively utilizing diverse function calls, and the tendency to hallucinate less-used geospatial libraries.
  To tackle these problems, we introduce GeoAgent, a new interactive framework designed to help LLMs handle geospatial data processing more effectively.
  GeoAgent pioneers the integration of a code interpreter, static analysis, and Retrieval-Augmented Generation (RAG) techniques within a Monte Carlo Tree Search (MCTS) algorithm, offering a novel approach to geospatial data processing.
  In addition, we contribute a new benchmark specifically designed to evaluate the LLM-based approach in geospatial tasks.
  This benchmark leverages a variety of Python libraries and includes both single-turn and multi-turn tasks such as data acquisition, data analysis, and visualization.
  By offering a comprehensive evaluation among diverse geospatial contexts, this benchmark sets a new standard for developing LLM-based approaches in geospatial data analysis tasks.
  Our findings suggest that relying solely on knowledge of LLM is insufficient for accurate geospatial task programming, which requires coherent multi-step processes and multiple function calls.
  Compared to the baseline LLMs, the proposed GeoAgent has demonstrated superior performance, yielding notable improvements in function calls and task completion. 
  In addition, these results offer valuable insights for the future development of LLM agents in automatic geospatial data analysis task programming.
\end{abstract}

\keywords{Code Generation, Agent, LLM, Benchmark, Geospatial Data Analysis}

\maketitle
\section{Introduction}
Large language models have demonstrated their potential to solve complex tasks in the geospatial domain \cite{Singh_2024_CVPR,10591792,Kuckreja_2024_CVPR,ZHANG2024103976,Xu2024RSAgentAR,Hu2023RSGPTAR,Bazi2024RSLLaVAAL,Zhan2024SkyEyeGPTUR,Guo2024RemoteSC}.
These initial efforts focus on enabling LLM to utilize external tools, plugins, etc, for addressing geospatial data processing tasks.
However, existing approaches are based on pre-defined low-level template-based prompts which often involve standalone third-party application programming interfaces (APIs) as foundational components for task completion.
Among these works, Remote Sensing ChatGPT \cite{Guo2024RemoteSC} and ChangeAgent \cite{10591792} focus on utilizing independent remote sensing (RS) vision model APIs while GEOGPT \cite{ZHANG2024103976} uses third-party API calls provided by the GIS system.
These APIs, particularly based on given systems, provide single-line API calls for a specific task without a deep understanding of the dependencies of different functionalities.
A more recent effort, GeoLLM-Engine \cite{Singh_2024_CVPR}, integrates several APIs into a sequential task on a real user interface platform.
This is closer to real-world geospatial data processing tasks including the multi-steps of online data acquisition, data analysis, and visualization in one system.
However, this work is still limited by the task-level APIs and lacks flexibility in complex and open-domain geospatial data analysis scenarios.
To tackle these challenges, more recent efforts in natural language processing (NLP), such as QwenAgent \cite{bai2023qwen}, DS1000 \cite{lai2023ds}, ODEX \cite{wang2022execution}, BigCodeBench \cite{zhuo2024bigcodebench} and CIBench \cite{zhang2024cibench}, aim to leverage the code generation capabilities of LLM for executing a wide range of open domain data analysis tasks, utilizing any available Python libraries.
These studies illustrate the potential of LLMs in solving open-domain data analysis tasks within code execution environments and offer promising directions for applying similar approaches to geospatial data analysis in open-domain settings.

Geospatial data analysis tasks \cite{wu2024geospatial}, consisting of data collection, data storage, data retrieval, data analysis, prediction, and visualization, present huge challenges for LLMs to follow complex instructions and accurately utilize specialized libraries and models.
The tasks require a comprehensive understanding of instructions, including the inter-relationships among various inputs and outputs, systematic task decomposition, and the application of specialized domain expertise.
Expert intervention is occasionally necessary for task decomposition and dynamic adjustments, which may involve assigning specific libraries.
Developing executable solutions requires invoking multiple function calls from various libraries, which is often challenging for LLMs, particularly when they are under-trained with these libraries. 
General data analysis work typically relies on existing benchmarks \cite{lai2023ds,wang2022execution,zhuo2024bigcodebench,zhang2024cibench}, which primarily involve frequently used APIs from popular Python libraries. 
The efficacy of API usage in these studies has been saturated by the recently released models accompanied by extensive online documentation and limited to single-turn question assessments.
These studies have a limited scope in geospatial data analysis, which often requires using several geospatial Python libraries in a given task with consecutive steps.
For less frequent APIs from geospatial Python libraries, LLMs tend to be troubled by the API hallucinations \cite{jain2024mitigating}, especially when under-trained on domain-specific knowledge.
This difficulty has been highlighted in recent studies \cite{zhuo2024bigcodebench}.
In NLP, retrieval-augmented generation (RAG) \cite{guu2020retrieval} has been proposed to incorporate domain-specific knowledge into task programming.
Retrieval-augmented approaches \cite{wang2024coderag} can enable models to base their predictions on external knowledge using various search techniques.
Simultaneously, advancements in new architectures and techniques have empowered models to process extensive sequences of tokens \cite{li2024long}.
However, effectively tackling specialized tasks, particularly in the geospatial domain, requires precise domain expertise, sequential multi-step reasoning, and iterative refinement guided by execution feedback.
This feedback is particularly crucial.
For example, the availability of online data may not be evident before task execution, highlighting the limitations of currently available general-purpose LLM coders.
In addition, incorporating sequential reasoning capabilities often relies on Monte Carlo Tree Search (MCTS), which has demonstrated robust reasoning capabilities in many tasks where LLMs tackle complex mathematical problems \cite{zhang2024accessing,liao2024mario}.

In addition to the standalone API call, most existing works \cite{Zhan2024SkyEyeGPTUR,Bazi2024RSLLaVAAL,Hu2023RSGPTAR,Xu2024RSAgentAR,Kuckreja_2024_CVPR} addressing geospatial data analysis tasks through large Vision-Language Models (VLMs) \cite{li2024vision} by integrating input data with task descriptions in a single operation.
Although both methods are more direct, they lack comprehensive analysis of individual results and do not facilitate the integration of various models' outputs and data resources for in-depth analysis.
Several key challenges arise when employing large models in practical geospatial tasks. 
First, understanding geospatial data is complex due to its various modalities, which carry richer information than RGB images.
The requirement of substantial model parameters in VLMs to effectively initialize this knowledge is often impractical. 
Second, invoking domain knowledge is essential for addressing interdisciplinary topics. 
For example, the physics models of specific RS data are embedded within online tutorials and geospatial libraries, which are typically inaccessible to current VLMs. 
Geospatial tasks such as moisture estimation from Synthetic Aperture Radar (SAR) data, involve several physics model functions, which cannot be achieved through a simple text prompt in VLMs. 
Third, compositional reasoning presents another challenge, as most geospatial tasks involve multiple components, such as image preprocessing and spatial relation analysis. 
Unlike prompting VLMs, LLM coder offers more adaptable solutions by integrating different models and facilitating comprehensive analyses. 
Lastly, online data access is a significant limitation; in remote sensing, vast datasets are stored in the cloud and accessed through APIs, but VLMs currently lack the ability to perform online data querying.

To address the aforementioned challenges, we introduce an LLM-based agent, dubbed GeoAgent, which integrates a code interpreter, static analysis, and RAG within an MCTS algorithm.
Additionally, we propose an evaluation benchmark encompassing diverse geospatial data analysis tasks, supplemented by an RAG library comprising geospatial library documents and solution examples.
This agent fulfills human requirements by translating a given task into executable Python code, utilizing dynamic task adjustment and refinement through the MCTS. 
During the code generation process, RAG invokes external knowledge into the LLM, particularly when employing less common geospatial libraries.
While RAG enhances task-specific coding proficiency by integrating corresponding library documents, it can negatively impact the performance of popular Python modules due to occasional irrelevant augmentations from suboptimal retrievers.
The process of dynamic refinement necessitates LLM iteratively improving code generation based on the feedback, especially when tasks involve multiple consecutive steps that build upon each other. 
This method identifies dependencies among subtasks and dynamically refines them using execution feedback within an MCTS framework, ensuring that each code segment is logically consistent and well-developed in relation to prior steps. 
The following outlines our key contributions:
\begin{itemize}
\item We introduce a novel LLM agent that integrates an external knowledge retriever, a code interpreter, and static analysis within MCTS, tailored for geospatial data analysis.
This integration enhances problem-solving, and logical capabilities in sequential task programming.
This agent operates within a Jupyter Notebook environment, allowing users to interact with the model iteratively, refining and optimizing task execution.
This ensures compliance with Python libraries and best practices, while API tokens for various libraries are securely managed on a remote server, reducing the burden on end-users to acquire multiple tokens.

\item We present GeoCode, an execution-based benchmark comprising over 18,000 single-turn and 1,356 multi-turn geospatial data analysis tasks, involving 2,313 function calls from 28 widely-used libraries across 8 task categories.
GeoCode features two distinct evaluation models: single-turn task evaluation, which assesses function call accuracy and individual task success, and multi-turn task evaluation, which focuses on task completion rates. 
The latter operates in two modes: automatic iterative refinement sequentially and human intervention for failed subtasks.

\item We evaluate various LLMs on the GeoCode benchmark. While general-purpose LLM coders may produce incomplete or incompatible workflows, GeoAgent achieves higher pass and task completion rates, demonstrating superior performance. 
This success is attributed to its effective handling of specialized Python libraries within an MCTS framework, guided by execution feedback. 
The study provides a more accurate assessment of LLM coders in geospatial data processing and points to a promising future for automatic geospatial data analysis.
\end{itemize}

The rest of this paper is organized as follows.  
Section 2 presents the related works on LLM-based code generation and LLM-empowered geospatial data analysis.   
Section 3 introduces the proposed method by describing the framework of GeoAgent, API retrieval, and dynamic refinement on MCTS.
The descriptions of the experimental setup, benchmark statistics, RAG library, LLM, and baseline settings as well as the evaluation metrics in Section 4.
Evaluation results obtained on function call performance, task-level performance, as well as discussion, are illustrated in Section 5. 
Finally, Section 6 concludes the paper.

\section{Related Works}
\subsection{LLM-based Code Generation} 
LLMs exhibit a strong ability to generate code based on the pre-training of large-volume code sets.
Although these LLM coders have proven highly effective in generating standalone functions, they face challenges dealing with interrelated tasks.
To address these challenges, researchers have proposed several strategies to enhance the programming capabilities of LLMs, including instruction tuning \cite{muennighoff2023octopack}, self-debugging \cite{chen2023teaching,jiang2024training}, and in-context learning \cite{li2023large,patel2023evaluating}.
More recent pre-trained LLM coders have leveraged instruction tuning further for repository-level coding task \cite{zhang2023repocoder,jimenez2023swe}.
Instruction-tuned LLMs generally excel at producing code snippets that are well-aligned with specific natural language (NL) task instructions.
Despite these improvements, LLMs still face difficulties in programming tasks requiring multi-step logic, particularly in integrating with data science processing pipelines.
Current efforts \cite{yao2022react,wang2024executable,xia2024aicodereval,zhang2024cibench} focus on boosting the LLMs' performance in data science applications and enabling code-based reasoning capabilities by invoking error traceback and analysis. 
For instance, the incorporation of code interpreters \cite{zhang2024cibench} and static analysis \cite{xia2024aicodereval} within the workflow is noteworthy.
For example, RepairAgent \cite{bouzenia2024repairagent} utilizes static analysis to provide LLM-based repair insights, while CoderGen \cite{xia2024aicodereval} employs static analysis to verify variables and statements. 
Additionally, $STALL^{+}$ \cite{liu2024stall+} uses static analysis to generate contextual information for iterative LLM-based code repairs.
These studies underscore the benefits of combining static analysis with LLMs for coding tasks.
In the context of execution feedback, CodeAct \cite{wang2024executable} executes and dynamically improves code generation through multi-turn interactions with a Python interpreter and feedback from static analysis.
The interpreter within this iterative framework is capable of managing symbolic computations, logical reasoning, and precise numerical operations, with its feedback contributing to the enhancement of loop consistency. 
In the realm of retrieval-augmented LLM coders \cite{patel2023evaluating,li2023large,wang2024coderag,jain2024mitigating}, linking LLMs to up-to-date external databases improves the precision of code generation,
particularly when leveraging less popular and continuously evolving libraries to address specific challenges.
However, we lack a geospatial RAG database to prompt programming for geospatial tasks.
Another important direction is to integrate standalone APIs from external tools into the processing loop.
For instance, Schick et al. propose ToolFormer \cite{schick2024toolformer}, which uses extra information to instruct LLMs on how to invoke existing APIs.
Zhang et al. subsequently developed ToolCoder \cite{zhang2023toolcoder} to finetune LLMs on API utilization during code generation. 
Although ToolCoder effectively handles the invocation of standalone APIs within code generation, it does not adequately address open-domain tasks involving open libraries.
Additionally, Liu et al. \cite{liu2024exploring} have identified that API hallucinations account for up to 15\% of all hallucinations in state-of-the-art LLM coders, particularly with less common ones. 

\subsection{Geospatial Analysis with Large Language Models} 
Researchers \cite{li2023autonomous,manvi2023geollm,zhang2024geogpt,lin2023geogalactica,deng2024k2,hu2023rsgpt,zhang2024earthgpt,bazi2024rs} have explored the incorporation of LLMs into the processing and analysis of geospatial data.
The initial efforts have focused on employing LLMs with text-based inputs such as Geographic Information System (GIS) tasks \cite{zhang2024geogpt,lin2023geogalactica,deng2024k2}. 
For instance, Li et al, \cite{li2023autonomous} have tested the spatial semantic reasoning ability of ChatGPT \cite{Ray2023ChatGPTAC} on geospatial tasks including toponym recognition, location description, and time series forecasting.
This study indicates that LLMs are capable of comprehending and parsing geospatial data processing and analytical tasks based solely on their intrinsic knowledge. 
While this approach leverages the spatial analytical potential of LLMs, it is constrained by the lack of direct access to visual information, which is often crucial in geospatial analysis. 
The preliminary attempt at bridging the gap between visual features and the semantic reasoning capabilities of LLMs is large VLMs \cite{hu2023rsgpt,zhang2024earthgpt,bazi2024rs,kuckreja2024geochat}, which incorporate LLMs into RS tasks, such as RS image captioning, Visual Question Answering (VQA), and visual grounding.
Pioneering work RSGPT \cite{hu2023rsgpt} built a high-quality human-annotated RS image captioning dataset that advances the development of VLMs in the RS domain.
Additionally, RSGPT introduced a GPT-based model specifically designed for RS image captioning and VQA.
To better leverage the ability of LLMs, RS-Llava \cite{bazi2024rs} create the RS-instructions dataset, a comprehensive benchmark dataset that integrates four diverse single-task datasets related to captioning and VQA in a Llava \cite{Liu2023ImprovedBW} framework. 
However, the textual outputs of LLMs often fall short of meeting users' precise expectations.
In addition to generating textual output, recent advancements in GeoChat \cite{kuckreja2024geochat} have incorporated a broader range of tasks within VLMs, encompassing referring expression, region captioning, image description, and VQA. 
Although these developments have yielded good results, they remain constrained to RGB images and are exclusively focused on computer vision tasks.

Despite the remarkable achievements of these models in various geospatial data analysis tasks, leveraging these models to address complex challenges across open-domain tasks remains a significant difficulty. 
The challenges associated with spatial analysis and multimodal data analysis present significant hurdles within the LLMs and VLMs paradigm.
Specifically, tasks within the geospatial data processing field inherently rely on the utilization of multiple professional tools, procedures, and multimodal data. 
Recent attempts \cite{ningautonomous,zhang2024geogpt} explore the potential of converting LLMs from a purely operational role to a decision-maker. 
Rather than directly addressing tasks, these approaches capitalize on the exceptional capabilities of LLMs in language comprehension and reasoning to assess requirements and select appropriate professional tools.
Subsequently, the tasks are carried out using these selected tools.
In this context, a prior initiative known as GEOGPT \cite{zhang2024geogpt} offers a viable framework for addressing GIS tasks.
GeoGPT integrates LLMs with established tools within the GIS field to tackle a range of geospatial challenges. 
Change-Agent \cite{10591792} uses an LLM to follow user instructions to call segmentation and captioning tools for the RS change detection task.
In addition, GeoLLM-Engine \cite{Singh_2024_CVPR} calls more geospatial API tools and external knowledge bases, enabling agents to handle complex geospatial tasks.
This work delineates the precise requirements for these tools, which must be pre-determined and wrapped into a task-level API.
This limits the application on open-domain tasks.
Furthermore, tool calling defeats on the integration of task-level APIs in a sequential task.

\begin{figure}[pt]
\centering
\includegraphics[width=\linewidth]{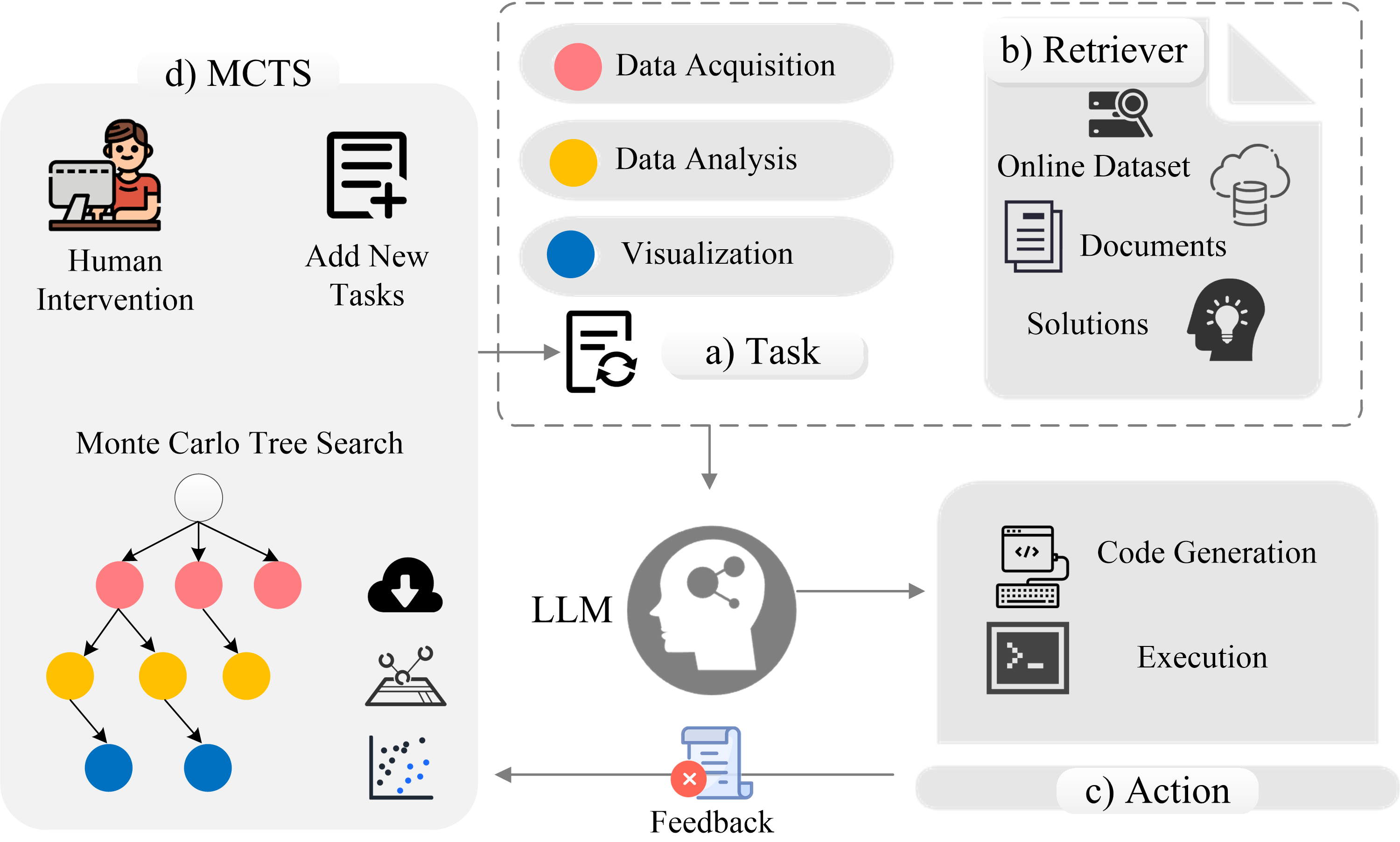}
\caption{GeoAgent: A geospatial data analysis task programming agent. This agent comprises four integral components: \textbf{b) Retriever}, which retrieves task-relevant items from provided document data, including Python library documents, online tutorials, and function-level solutions; \textbf{c) Action}, which is an execution environment integrated with a code interpreter and static analysis, provides feedback on generated code and suggests potential fixes; \textbf{d) MCTS}, explores and evaluates multiple possible code candidates to optimize the selection of the most promising solution at each step through iterative adjustment and refinement; and \textbf{LLMs}, function as both the code generator and the intelligent reasoner to propose code solutions and iteratively diagnose and fix errors.}
\label{fig1}
\end{figure}

\section{Methodology}
\subsection{Framework} 
In this section, we present GeoAgent, an LLM agent designed for the processing and analysis of geospatial data, tailored to meet the needs of researchers. 
The overall architecture of GeoAgent is illustrated in Fig. \ref{fig1}, which highlights two components:
1) Task programming: GeoAgent starts by leveraging parameterized knowledge of LLMs to generate code based on task instructions. When a specific Python library is mentioned, GeoAgent retrieves relevant APIs from the RAG database (see Section 3.2) and generates codes conditioned on the retrieved items.
2) Task refinement on MCTS: GeoAgent executes and collects execution feedback within the MCTS framework, where MCTS iteratively explores and evaluates multiple code candidates to optimize the selection of the most promising solution, while LLMs serve as the intelligent reasoner, diagnosing and fixing errors.
Notably, RAG enables GeoAgent to interface with various Python libraries of different geoscience domains. 
This capability allows GeoAgent to incorporate external knowledge for specialized domain tasks.
While RAG provides essential information for specific tasks, GeoAgent still requires dynamic adjustments, such as improving initial prompts and correcting any failed generated code.
In this study, we integrate LLMs with MCTS to facilitate dynamic adjustments (see Section 3.3) during task programming. 
Our self-refinement algorithm employs an iterative refinement process within a search tree framework, enabling a step-by-step improvement.

\subsection{API Retrieval}
In this work, the RAG database incorporates diverse Python library documents, online tutorials, and function-level solutions accessible to LLMs. 
GeoAgent utilizes the external RAG database to identify relevant functions based on the provided task instructions and previous codes.
Once suitable solutions and best-fit functions are retrieved, GeoAgent leverages the power of LLMs to effectively transform given instructions into executable codes.
In addition, when the generated code fails to call specific functions due to API hallucinations, GeoAgent retrieves their usage details from the specified Python library.
For a given task $t_i$, GeoAgent employs the Retriever module to extract relevant function information and usage examples from the RAG database.
GeoAgent utilizes the off-the-shelf embedding model as the selector.
The retrieval process involves three steps: First, compute the embedding features of the given task description $t_i$ and the retrieved function documents. 
Second, calculate the cosine similarity between these embeddings. 
Finally, the $k$ most relevant functions are returned, denoted as $A_i = \langle Func., Usage \rangle$, providing contextual information of the given task $t_i$ to LLMs.

\begin{figure}[pt]
\centering
\includegraphics[width=\linewidth]{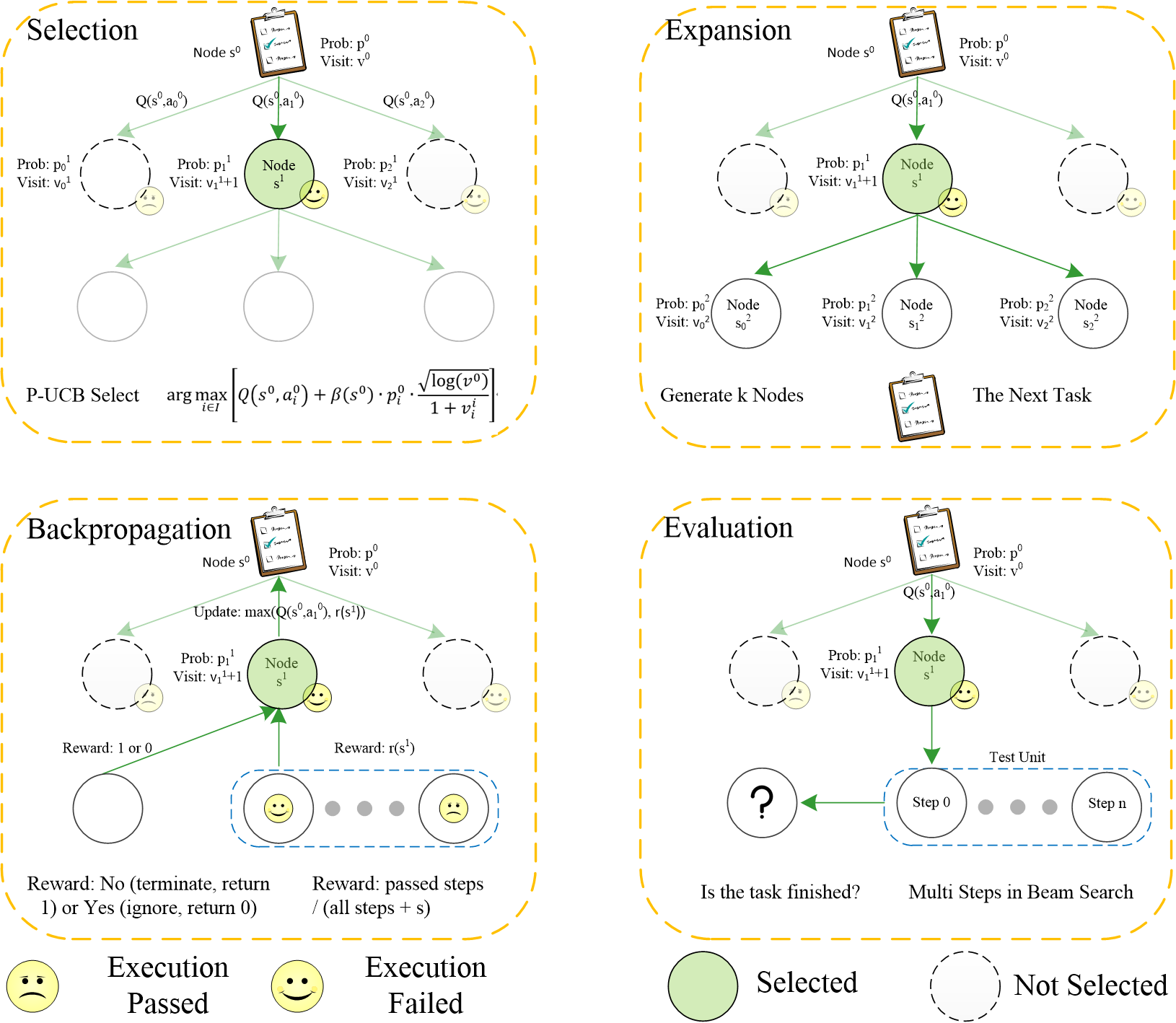}
\caption{Illustration of using the Monte Carlo tree search algorithm in geospatial data analysis task programming. }
\label{fig2}
\end{figure}

\subsection{Dynamic Refinement on MCTS}
The GeoAgent employs an MCTS framework \cite{zhang2023planning} to iteratively refine each subtask within a sequential task.
MCTS utilizes a tree structure where nodes represent states $s$ and edges denote actions $a$.
The algorithm begins at the root node $s^0$ and explores the state space to identify the terminal state $s^n$ with the highest reward $r(s^n)$.
Each node comprises the following components: visited count $v$: the visited times of each node; probability $p$: derived from LLMs; state-action value $Q(s, a)$: the maximum reward obtained by taking action $a$ from state $s$.
The reward $r$ of state $s$ is defined as the proportion of successfully passed steps.
This algorithm prioritizes visiting nodes with higher reward values, as these indicate high-quality generation. 
Additionally, it explores nodes with fewer visits to ensure under-explored areas of the state space are adequately investigated.

In this framework, GeoAgent first transforms NL instructions into executable codes and then executes them.
Nevertheless, geospatial tasks involve extensive operations and intricate workflows, making it challenging to generate coherent multi-step code in one attempt.
This complexity necessitates a comprehensive understanding of the context, as well as continuous feedback and adjustments of each step.
To tackle this challenge, GeoAgent employs beam search combined with execution filtering in child node sampling and prompt updating throughout the MCTS expansion process.
The entire process of MCTS is illustrated in Fig. \ref{fig2}.

The process begins at the root node, which initiates from the first subtask and progresses sequentially through previous ones.
During the selection phase, the appropriate branch starting from root node $s^0$ is chosen utilizing the probabilistic Upper Confidence Bound ($p-UCB$) algorithm:
\begin{equation}
p-UCB=\arg \max _{i \in I}\left[Q\left(s^0, a_i^0\right)+\beta\left(s^0\right) \cdot p_i^0 \cdot \frac{\sqrt{\log \left(v^0\right)}}{1+v_i^i}\right] ,
\end{equation}
where $I$ is the collection of children nodes; $\beta(s^0)$ is the weight for exploration, which depends on the number of visits ($v^0$) and constants $c_{base}$ and $c$.
\begin{equation}
\beta(s^0) = \log\left(\frac{v^0 + c_{base} + 1}{c_{base}}\right) + c .
\end{equation}
The $p-UCB$ algorithm incorporates an exploration term $\beta(s^0)$, weighted by the probability $p$ of the LLM-determined sequence score, where an increased value of $c$ promotes greater exploration.
For instance, GeoAgent starts from the root node $s^0$ and selects subtrees recursively until it reaches a node that has not been expanded.
In this process, $p-UCB$ balances the exploration between known-to-be-good states and less visited states.
In the expansion phase, after selecting the initial node, potential codes for subsequent steps are generated and added as new nodes to the child node list until the next subtask is reached.
In this process, we sample $n$ output sequences generated from the prompt. 
From these output sequences, only top $k=3$ sequences are returned. 
These $k$ output sequences ($s_0^2, s_1^2, s_2^2$) are subsequently added to the children list of the current code $s^1$.
For each child node, we define their reward value $Q$ as $0$ if they are non-terminal reasoning steps and $Q$ value as $-1$ if they failed the evaluation test.

During the evaluation phase, the selected node $s^1$ must be assessed, despite the node potentially representing only a partial program.
The quality of a partial program cannot be directly evaluated, as its ability to solve the task and adhere to instructions remains uncertain.
To address this, the LLM does a look-ahead search from the current node by generating a multi-step code (test unit) in one attempt.
Specifically, given the code of the current node, we first collect code candidates generated by LLMs. 
The generated code candidate is then concatenated with the code from previous nodes to verify whether the combined code can pass the execution test, ensuring that no parsing or completion errors are detected by the static analyzer and code interpreter. 
Finally, the reward of the current node is calculated based on the multi-step code, measured as the ratio of executable steps to all steps. 
Additionally, a reward of 1 is assigned to the current node when the multi-step code completely overlaps with the current node.
This reward $r(s^1)$ is then backpropagated through the tree, updating the values of its ancestor nodes accordingly.

\begin{figure}[pt]
\centering
\begin{minipage}{0.48\linewidth}
    \centering
    \includegraphics[width=\linewidth]{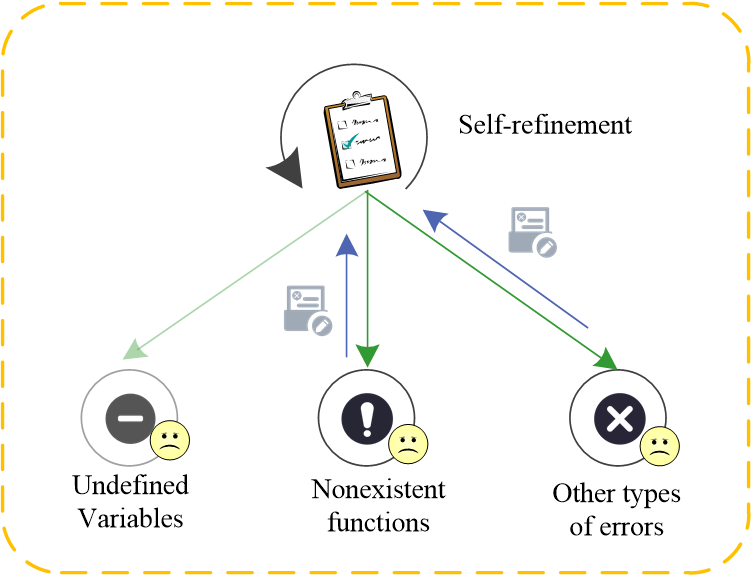}
    \caption{The self-refinement algorithm in MCTS.}
    \label{fig3}
\end{minipage}\hfill
\begin{minipage}{0.48\linewidth}
    \centering
    \includegraphics[width=\linewidth]{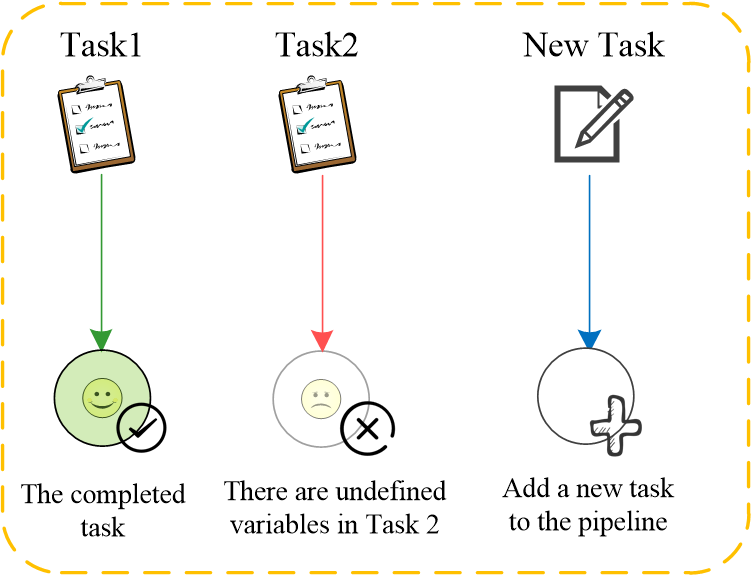}
    \caption{Addition of new tasks to the task pool in MCTS.}
    \label{fig4}
\end{minipage}
\end{figure}

The proposed MCTS framework incorporates a comprehensive error traceback and analysis mechanism to dynamically refine each subtask. 
This mechanism, depicted in Fig. \ref{fig3} and Fig. \ref{fig4}, operates by passing the initially generated code within a code interpreter and static analysis tools to assess its functionality.
It is common for LLMs to invoke undefined variables. 
Our methodology addresses this issue by removing the subtree that uses undefined variables and inserting a new task into the workflow to define the variable before its use in subsequent tasks, as shown in Fig. \ref{fig4}.
If incorrect function calls are present within the current nodes, we utilize static analysis tools, such as Python Jedi \cite{jedi}, to retrieve all accessible APIs (e.g., method names, variable names, and parameter names) of the relevant object. 
This information is then incorporated into the prompt, guiding the LLM to select the appropriate option and regenerate the current node accordingly.
For other error types, the interpreter captures an error traceback, providing a detailed record of the execution path that led to the failure. 
This traceback is then analyzed by the LLM, which identifies the point of failure and suggests potential fixes, whether they are syntactic errors, logical errors, or issues with Python libraries.
The failed code snippet, alongside the suggested corrections and the original instructions, is reintroduced into the LLM, which subsequently generates a revised version.
This process continues until the code successfully executes or the maximum number of attempts is reached.

However, even with the integration of static analysis and execution feedback into the prompting process, ensuring the success of task completion remains challenging due to the inherent non-determinism and limited interpretability of LLM inference.
To mitigate the risk of an endless loop, a maximum number of attempts for each task is established.
The capacity of the framework to learn from error tracebacks and real-time feedback is a distinctive feature that differentiates GeoAgent from traditional code generation processes. 
The self-debugging in MCTS can significantly reduce the time and effort required for scientists to conduct geospatial task programming.
Meanwhile, manual editing is also introduced to enhance the completeness and accuracy of sequential tasks.
If the problem goes unsolved, GeoAgent generates a report detailing the task, code, and encountered errors.
At this stage, human intervention is introduced, allowing for manual edits to failed subtasks based on the printed error information. 
In subsequent attempts, these human modifications are incorporated as contextual inputs.
Additionally, GeoAgent offers functionalities to utilize chat histories for improving unsuccessful code generations and include them in the new round. 
By incorporating both self-debugging and manual editing strategies within the MCTS framework, GeoAgent aims to produce code that is not only correct but also efficient and maintainable, reflecting best practices and external knowledge relevant to the task.

\section{Experimental Setup}

\subsection{Benchmark Construction} 
In this section, we outline the process of constructing the benchmark, which involves three primary steps: code collection, code refactoring, and instruction generation.
Unlike the short code exercise, constructing a high-quality, execution-based benchmark that focuses on geospatial programming tasks is non-trivial. 
The primary challenge lies in the difficulty of naturally sourcing self-contained geospatial programming tasks with detailed instructions. 
While many GitHub repositories contain realistic source code, the scripts inside these repositories often require cross-file functions.
Additionally, the diversity of geospatial task programming scenarios further complicates this effort. 
Meanwhile, to assess the performance of GeoAgent on geospatial tasks, the evaluation benchmark must be able to reply to several key questions: 1) Can LLMs follow complex geospatial task instructions? 2) Can LLMs-based geospatial task programming benefit from external knowledge? 3) Can LLMs enhance task completion by integrating MCTS?

This work specifically focuses on Python code due to its widespread use in geospatial data analysis.
Whereas the methodology employed in this task is adaptable and can be expanded with more programming languages in future benchmarks. 
GeoAgent gathers code snippets that utilize geospatial Python libraries and creates corresponding task descriptions for each code segment.
Typically, repository-level code is used to generate function-level programming tasks, but this approach often necessitates cross-file information, which can be challenging to isolate into standalone scripts.
Given the limited resources for geospatial data processing and analysis within Python scripts, GeoAgent employs tutorials from various Python libraries and leverages LLMs to construct code-instruction pairs.
Notably, we split GeoCode into two subsets: Google Earth engine library-based tasks (GeoCode-GEE) and the other library-based tasks (GeoCode-Other).

\subsection{Benchmark Statistics} 
The proposed GeoCode benchmark, summarized in Table \ref{tab1}, encompasses 28 widely-used Python libraries across 8 key domains. 
These libraries are extensively utilized in geospatial data analysis tasks, and the code snippets in GeoCode often involve combined function calls from multiple libraries, thereby requiring considerable compositional reasoning ability.
Additionally, Table \ref{tab2} compares GeoCode with executable Python programming benchmarks, highlighting the libraries and function calls referenced in these benchmarks.
Notably, GeoCode includes 18148 single-turn tasks and 1356 multi-turn tasks as well as 2313 function calls from 28 external libraries, reflecting a broader diversity compared to other benchmarks. 
This indicates that GeoCode offers diverse task prompts, which involve complex instructions and require the development of solution code with intricate implementation logic.

\begin{table}[pt]
\caption{Illustration of Python libraries in GeoCode. This table categorizes Python libraries used in GeoCode according to their respective domains. Each domain is associated with specific tasks.}
\renewcommand\tabcolsep{15pt}
\centering
\label{tab1}
\begin{tabular}{ccc}
\hline
Domain              & Library  \\ \hline
Data Acquisition \& Preparation & \begin{tabular}[c]{@{}c@{}}earthengine-api, cubo, pystac, GOES-2-Go, meteostat,\\ pystac\_client, pytesmo, planetary\_computer\end{tabular} \\
Raster Processing   & earthengine-api, eemont, geetools, GeoUtils, wxee, xarray-spatial \\
Vector Processing   & GemGIS, GeoPandas, GeoUtils \\
3D analysis         & Gempy \\
Machine Learning    &scikit-eo, Verde \\
Deep Learning       & segment-geospatial, srai  \\
Specific Alogorithm & geeet, gstools, sen2nbar, pylandtemp, eradiate, spectramap \\
Visualization       & geemap, leafmap, Lonboard \\ \hline
\end{tabular}
\end{table}

\begin{table}[pt]
\caption{Python programming benchmark statistics: analysis by external library usage, function call frequency, task count, and task type.}
\label{tab2}
\renewcommand\tabcolsep{12pt}
\centering
\begin{tabular}{ccccc}
\hline
Benchmark    & Ext. Library & Function Call & Single-Turn Tasks & Multi-turn Tasks\\ \hline
DS-1000      & 14           & 540           & 1000   & 0         \\
ODEX         & 13           & 190           & 439    & 0         \\
BigCodeBench & 62           & 877           & 1140   & 0         \\
CIBench      & 11           & 171           & 469    & 73        \\ \hline
GeoCode      & 28           & 2313          & 18148  & 1356        \\ \hline
\end{tabular}
\end{table}

\subsection{RAG Library}
Generating code solely with LLMs based on parametric knowledge is challenging, as it cannot keep pace with evolving public libraries and datasets.
In this work, we integrate three key resources to enable models to retrieve relevant contexts: specific solutions, online tutorials, and library documents.
1) Specific solutions: We compiled a document containing basic solutions to specific functionalities consisting of natural language descriptions and code solutions.
2) Online tutorials: We aggregated tutorials from various websites, each page comprising code snippets and textual explanations on topics ranging from change detection to computer vision.
3) Library documents: We collected official documentation for all Python libraries shown in Table \ref{tab1}, which is particularly valuable for libraries with scarce online tutorials.
However, current retrieval systems struggle to source useful contexts, especially when shared vocabulary is limited.
And, code generators often fail to improve performance when integrating irrelevant contexts.
To address these challenges, we implement RAG enhanced with metadata filters.

\subsection{Experimental Settings}
In this work, we select Llama 3.1 (8B) \cite{Dubey2024TheL3} as the LLM to support geospatial data analysis programming.
Additionally, we consider CodeGemma 2 (7B) \cite{Zhao2024CodeGemmaOC} and Phi3.5 mini (3.8B) \cite{Abdin2024Phi3TR}, Qwen2 (7B) \cite{Yang2024Qwen2TR} in our analysis of the impact of different model sizes and specialized LLMs.
Inference is performed using a single NVIDIA GeForce RTX 4090 GPU, encompassing both the initial stage of instruction generation and the subsequent stage of code generation.
The parameter configurations include setting the top-p value to 0.9, the temperature parameter to 0.6, and a maximum token limit of 2048.
The window size is set to at least 32k tokens, thereby supporting the retrieval of extensive contexts. 
For the Retriever modules, we employ the BBAI-embedding-001 model to generate high-quality embeddings for both text and code, enabling efficient similarity computations and retrieval processes.

\subsection{Metrics} 
In this section, we introduce the evaluation metrics utilized for assessing the function call and task completion.
To evaluate the performance of function calls within generated code, we employ evaluation metrics typically used in multilabel classification tasks: Precision, Accuracy, Recall, F1 score, and Hamming distance. 
We extract the functions referenced in the generated code as predicted labels. Since the function list serves as a set of labels, predictions may be entirely correct, partially correct, or entirely incorrect.
This aligns the evaluation with the exampled-based multilabel classification \cite{10.1007/978-3-319-16483-0_54}, where partial correctness is taken into consideration.
We utilize an example-based strategy, which involves calculating the average difference between predicted and actual labels for each task and then averaging these differences across all tasks in the test set.
Let $T$ be a function call evaluation dataset consisting $n$ tasks ($X_i, Y_i$), $1 \leq I \leq n$ and $x_i \in X, y_i \in Y=\{0, k\}^k$, with $k$ classes.
Let $h$ denote the LLM and $Z_i = h(x_i) = \{0, 1\}^k$ represent the set of label memberships predicted by the LLM for the task $x_i$.
Accuracy for each instance is defined as the proportion of correctly predicted labels to the total labels (both predicted and actual ones) for one instance. The overall accuracy is the average of these proportions across all instances:
\begin{equation}
A=\frac{1}{n} \sum_{i=1}^n \frac{\left|Y_i \cap Z_i\right|}{\left|Y_i \cup Z_i\right|}
\end{equation}
Recall is defined as the proportion of correctly predicted labels to the total number of actual labels, averaged across all instances:
\begin{equation}
R=\frac{1}{n} \sum_{i=1}^n \frac{\left|Y_i \cap Z_i\right|}{\left|Z_i\right|}
\end{equation}
Precision is defined as the proportion of correctly predicted labels to the total number of predicted labels, averaged across all instances:
\begin{equation}
P=\frac{1}{n} \sum_{i=1}^n \frac{\left|Y_i \cap Z_i\right|}{\left|Y_i\right|}
\end{equation}
The F1 score, which is the harmonic mean of precision and recall, is calculated as follows:
\begin{equation}
F_1=\frac{1}{n} \sum_{i=1}^n \frac{2\left|Y_i \cap Z_i\right|}{\left|Y_i\right|+\left|Z_i\right|}
\end{equation}
Hamming loss measures the frequency with which a label is incorrectly predicted or a relevant label is missed. It is normalized by the total number of classes and examples.
\begin{equation}
\text { Hamming loss}=\frac{1}{k n} \sum_{i=1}^n \sum_{l=1}^k\left[I\left(l \in Z_i \wedge l \notin Y_i\right)+I\left(l \notin Z_i \wedge l \in Y_i\right)\right]
\end{equation}
where $I$ is the indicator function and a smaller Hamming loss indicates better performance.
To analyze the impact of RAG on function calls, we compare the evaluation metrics under conditions where zero and three retrieved items are used, respectively.

For assessing performance in code completion, metrics such as pass rate and task completion rate (the number of successfully executed steps achieved divided by the total number of steps) are considered, where pass rate is mainly for single-turn task evaluation and task completion rate being the primary measure of multi-turn tasks.
For example, if a task consists of 10 steps, we provide the model with 10 prompts, allowing it to generate 10 code blocks sequentially. 
The completion rate is then calculated based on the consecutive successful executions, with a 50\% completion rate if the first five steps are passed.
Furthermore, the pass@k metric (with k=1 in this case) is a widely recognized evaluation measure that assesses the execution correctness of single-turn tasks. For multi-turn tasks, sequential tasks were converted into single-turn tasks to facilitate this evaluation.

\section{Evaluation}
In this section, we present the evaluation results for all benchmarks along with the proposed GeoCode-GEE and GeoCode-Others. 
GeoCode-GEE is focused exclusively on tasks within the GEE environment, while GeoCode-Others requires the use of multiple Python libraries. 
Additionally, we include benchmarks relevant to general data science tasks, such as DS-1000, ODEX, BigCodeBench, and CIBench.
To address the questions posed in the benchmark construction, we first analyze qualitative function calls using Precision, Recall, F1 score, and Hamming loss metrics. 
Subsequently, we assess GeoAgent's performance in task completion using the pass rate and task completion metrics.

\begin{table}[pt]
\caption{Comparison of function call performance with zero (@0) and three (@3) retrieval items across all benchmarks.}
\label{tab3}
\renewcommand\tabcolsep{8pt}
\centering
\begin{tabular}{ccccccccccc}
\hline
\multirow{2}{*}{} & \multicolumn{2}{c}{Accuracy} & \multicolumn{2}{c}{Recall} & \multicolumn{2}{c}{Precision} & \multicolumn{2}{c}{F1} & \multicolumn{2}{c}{Hamming Loss} \\ \cline{2-11} 
                  & @0            & @3           & @0           & @3          & @0            & @3            & @0         & @3        & @0              & @3             \\ \hline
DS-1000           & 0.379         & 0.395        & 0.423        & 0.443       & 0.495         & 0.523         & 0.448      & 0.472     & 0.153           & 0.153          \\
ODEX              & 0.489         & 0.528        & 0.508        & 0.549       & 0.512         & 0.555         & 0.509      & 0.551     & 0.277           & 0.264          \\
BigCodeBench      & 0.369         & 0.611        & 0.411        & 0.675       & 0.524         & 0.805         & 0.449      & 0.724     & 0.108           & 0.101          \\
CIBench           & 0.543         & 0.571        & 0.604        & 0.634       & 0.658         & 0.713         & 0.621      & 0.662     & 0.139           & 0.138          \\ \hline
GeoCode-GEE       & 0.604         & 0.535        & 0.640        & 0.562       & 0.656         & 0.581         & 0.645      & 0.568     & 0.199           & 0.216          \\
GeoCode-Others    & 0.610         & 0.670        & 0.651        & 0.708       & 0.673         & 0.729         & 0.659      & 0.716     & 0.190           & 0.170          \\ \hline
\end{tabular}
\end{table}

\subsection{Function Call Performance} 
We first evaluate the function call performance considering the results obtained using Llama3.1 and RAG powered-Llama3.1.
All metrics are reported under settings that involve zero (@0) and three (@3) most relevant retrieval items.
As one can see in Table \ref{tab3}, most benchmarks benefit from the inclusion of retrieved function documents, leading to enhanced performance on all metrics.
The most significant improvement is observed on BigCodeBench, where the result of @3 achieves a 0.275 increase in F1 score and a 0.007 reduction in hamming loss.
This can be attributed to the less popular Python libraries being under-trained in Llama3.1, where the function definitions and usage examples from library documentation can substantially improve code generation.
Conversely, the improvement in F1 score for other benchmarks is minimal, with GeoCode-Others showing an increase of around 0.06, CIbench and ODEX seeing a rise of around 0.04, and DS1000 exhibiting almost no gain.
This limited improvement is likely due to DS1000 predominantly involving popular Python libraries, which are already saturated in the recently released Llama3.1 model. 
Notably, the performance on GeoCode-GEE shows a decline of including RAG.
This decline is likely because the GEE Python library is overly represented in the Llama3.1 training data, leading to saturation while the RAG includes additional noise in context and harms the generation.
This suggests that LLMs may not be sufficiently robust when dealing with excessive context, resulting in undesired behavior when processing varied context inputs.
It is important to note that most geospatial Python libraries are unlikely to become saturated in future LLM releases, as there are limited online documents available for further training. 
This highlights the importance of incorporating library documents in the context of LLMs to ensure the accurate invocation of best-fit functions. 
Additionally, retrieval accuracy plays a crucial role in the performance of function calls, necessitating further exploration of improved RAG algorithms.
In this work, we opt to implement RAG only when the library is mentioned, aiming to mitigate the impact of irrelevant retrieved items.

\subsection{Task level performance}
To assess GeoAgent's performance in task-oriented code generation, we conducted experiments involving 100 single-turn tasks of each benchmark and 10 sequential tasks of three multi-turn task benchmarks, given our limited computation resources. 
We first evaluated the pass rate (pass@1) of single-turn tasks under four different LLMs (i.e. Llama3.1, CodeGemma, Phi3.5-mini, Qwen2) using LLM alone and GeoAgent, respectively.
For single-turn task evaluation (as shown in Table \ref{tab4}), the code generated by Llama3.1 empowered GeoAgent seems troubled by the noise introduced in the multiple-step inference. 
Across all benchmarks except DS1000, it only increases the pass rate of GeoCode-GEE by 13\% and of ODEX by 17\% compared with the use of baseline Llama3.1 alone.
Notably, this noise can be avoided by adding an additional step in the loop.
However, if we consider all four different LLMs over all benchmarks, the LLMs alone have a lower pass rate, while GeoAgent increases the pass rate on most benchmarks.
This is because LLMs sometimes fail by calling incorrect or non-existent functions. 
GeoAgent addresses this issue by allowing multiple attempts to solve a task when the initial attempt fails.
Specially for GeoCode-GEE, GeoAgent achieves a 13\% improvement in pass rate for Llama3.1, a 6\% improvement for CodeGemma, and a 16\% improvement for Phi3.5-mini as well as an improvement of 20\% for Qwen2. 
Among the different LLMs, CodeGemma demonstrates the highest performance on the GeoCode benchmark, where achieved the pass@1 by 86\% on GeoCode-GEE and 59\% on GeoCode-Others.
This indicates a clear correlation between code instruction tuning and the performance of LLMs, with code post-training generally achieving better results. 
Among all benchmarks, DS-1000 shows a much lower pass rate overall LLMs and GeoAgents.
This issue is due to the absence of necessary code in the prompt, which leads to failed execution.
However, GeoAgent still improved its performance.
As for ODEX, BigCodeBench, and CiBench, they show less pass rate improvement than the proposed GeoCode.
Furthermore, the performance on GeoCode-GEE outperforms GeoCode-Others across all LLMs and GeoAgents, indicating that multi-library-based task programming is more challenging than single-library-based tasks. 
Whereas the pass rate alone does not fully capture task-level performance.
Hence, we present the F1 score of function calling of single-turn tasks under four different LLMs and GeoAgents (Table 5).
As one can see, GeoAgent improves function call performance compared to the vanilla LLM. However, we observed some declines in a few cases, particularly where the vanilla LLM had already achieved high performance.

\begin{table}[pt]
\caption{Code generation pass rate (Pass@1) of the Llama3.1 (8B), CodeGemma (7B), Phi3.5 mini (3.8B) and Qwen 2 (7B) on all benchmarks.}
\label{tab4}
\renewcommand\tabcolsep{8pt}
\centering
\begin{tabular}{ccccccccc}
\hline
               & \multicolumn{2}{c}{Llama3.1 (8B)} & \multicolumn{2}{c}{CodeGemma (7B)} & \multicolumn{2}{c}{Phi3.5 mini (3.8B)} & \multicolumn{2}{c}{Qwen 2 (7B)} \\ \cline{2-9} 
               & LLM & GeoAgent & LLM & GeoAgent & LLM  & GeoAgent & LLM  & GeoAgent \\ \hline
DS-1000        & 0.05 & 0.34 & 0.10 & 0.19 & 0.03 & 0.06 & 0.15 & 0.18 \\
ODEX           & 0.74 & 0.91 & 0.78 & 0.87 & 0.84 & 0.91 & 0.84 & 0.94 \\
BigCodeBench   & 0.67 & 0.61 & 0.82 & 0.82 & 0.94 & 0.91 & 0.87 & 0.84 \\
CIBench        & 0.93 & 0.92 & 0.93 & 0.96 & 0.94 & 0.99 & 0.96 & 0.93 \\ \hline
GeoCode-GEE    & 0.76 & 0.89 & 0.86 & 0.92 & 0.66 & 0.82 & 0.61 & 0.81 \\
GeoCode-Others & 0.45 & 0.40 & 0.58 & 0.76 & 0.50 & 0.55 & 0.39 & 0.61 \\ \hline
\end{tabular}
\end{table}

\begin{table}[pt]
\caption{Function call performance (F1 score) of the Llama3.1 (8B), CodeGemma (7B), Phi3.5 mini (3.8B) and Qwen 2 (7B) on all benchmarks.}
\label{tab5}
\renewcommand\tabcolsep{8pt}
\centering
\begin{tabular}{ccccccccc}
\hline
               & \multicolumn{2}{c}{LLama3.1 (8B)} & \multicolumn{2}{c}{CodeGemma (7B)} & \multicolumn{2}{c}{Phi3.5 mini (3.8B)} & \multicolumn{2}{c}{Qwen 2 (7B)} \\ \cline{2-9} 
               & LLM & GeoAgent & LLM & GeoAgent & LLM  & GeoAgent & LLM  & GeoAgent \\ \hline
DS-1000        & 0.83 & 0.86 & 0.75 & 0.75 & 0.80 & 0.80 & 0.79 & 0.79 \\
ODEX           & 0.53 & 0.66 & 0.54 & 0.56 & 0.55 & 0.58 & 0.55 & 0.56 \\
BigCodeBench   & 0.77 & 0.80 & 0.72 & 0.74 & 0.79 & 0.80 & 0.69 & 0.68 \\
CIBench        & 0.80 & 0.82 & 0.76 & 0.76 & 0.83 & 0.83 & 0.78 & 0.79 \\ \hline
GeoCode-GEE    & 0.78 & 0.86 & 0.75 & 0.71 & 0.70 & 0.77 & 0.76 & 0.74 \\
GeoCode-Others & 0.66 & 0.69 & 0.69 & 0.70 & 0.66 & 0.72 & 0.63 & 0.64 \\ \hline
\end{tabular}
\end{table}

\begin{table}[pt]
\caption{Code generation complete rate (Complete@1) of the Llama3.1, GeoAgent under modes of self-debugging (Aut.) and Human Intervention (Hum.) across all benchmarks. $idx$ refers to the task index, and $steps$ denotes the total number of task steps.}
\label{tab6}
\renewcommand\tabcolsep{7pt}
\centering
\begin{tabular}{ccccccc}
\hline
complete@1                & idx & steps & \multicolumn{1}{c}{Aut. Llama3.1} & \multicolumn{1}{c}{Hum. Llama3.1} & \multicolumn{1}{c}{Aut. GeoAgent} & \multicolumn{1}{c}{Hum. GeoAgent} \\ \hline
\multirow{10}{*}{CIBench} & 0 & 6 &0.50&0.66&0.17&0.83\\
                          & 1 & 6 &0.00&0.50&1.00& - \\
                          & 2 & 7 &1.00& - &1.00& - \\
                          & 3 & 7 &1.00& - &1.00& - \\
                          & 4 & 6 &1.00& - &1.00& - \\
                          & 5 & 6 &0.33&0.66&1.00& - \\
                          & 6 & 6 &1.00& - &1.00& - \\
                          & 7 & 7 &1.00& - &1.00& - \\
                          & 8 & 7 &0.14&0.86&0.43&0.43\\
                          & 9 & 7 &1.00& - &1.00& - \\ \hline
\multirow{10}{*}{GeoCode-GEE}  & 0 & 16 &0.19&0.81&0.19&0.88\\
                          & 1 & 8 &0.00&0.75&0.63&0.88\\
                          & 2 & 13 &0.08&0.69&0.23&0.85\\
                          & 3 & 22 &0.27&0.82&0.05&0.91\\
                          & 4 & 12 &1.00& - &0.42&0.83\\
                          & 5 & 7 &0.00&0.71&0.14&0.86\\
                          & 6 & 11 &0.27&0.73&0.09&0.82\\
                          & 7 & 8 &0.50&0.88&1.00& - \\
                          & 8 & 11 &0.55&0.73&0.45&0.82\\
                          & 9 & 9 &0.00&0.33&0.55&0.55\\ \hline
\multirow{10}{*}{GeoCode-Others}& 0 & 4 &0.00&0.50&1.00& - \\
                          & 1 & 6 &0.00&0.00&0.17&0.50\\
                          & 2 & 9 &0.00&0.78&0.33&0.78\\
                          & 3 & 7 &0.29&0.71&0.71&0.71\\
                          & 4 & 8 &0.13&0.25&0.25&0.25\\
                          & 5 & 4 &0.25&0.50&0.25&0.75\\
                          & 6 & 8 &0.00&0.00&0.00&0.00\\
                          & 7 & 6 &0.00&0.67&1.00& - \\
                          & 8 & 9 &0.00&0.67&1.00& - \\
                          & 9 & 7 &0.00&0.57&0.71&0.86\\ \hline
\end{tabular}
\end{table}

For sequential task evaluation (Table \ref{tab6}), we assess the completion rates of all benchmarks under self-debugging and human intervention modes, respectively.
When a problem remains unresolved, human intervention is introduced to advance the task to the next step.
Unlike single-turn tasks, sequential tasks challenge the reasoning capabilities of LLMs when generating code step by step.
Consistent with single-turn task evaluations, the GeoAgent achieves a higher completion rate on all benchmarks under both automatic and human intervention modes.
Compared with the vanilla Llama3.1, GeoAgent brings an improvement of complete rate by 20\% and 6\% on CIBench, 9\% and 10\% on GeoCode-GEE, 48\% and 22\% on GeoCode-Others under self-debugging and human intervention modes, respectively.
The possible reason is that the generated code using Llama3.1 alone often suffers from the use of undefined variables, leading to execution failures.
In such cases, dynamic adjustment of GeoAgent refactors undefined variables into new subtasks and updates failed attempts in the current task loop.
Compared with the automatic mode, human intervention brings an improvement of complete rate by 20\% and 7\% on CIBench, 46\% and 47\% on GeoCode-GEE, 40\% and 14\% on GeoCode-Others under LLMs alone and GeoAgent, respectively.
Across both self-debugging and human intervention modes, almost all cases perform better with the assistance of human expertise.
Whereas few hard cases still keep the same completion rate even with human intervention, such as the sixth task in GeoCode-Others.
These observations suggest that LLMs can achieve better results with human interaction, indicating a promising avenue for integrating LLMs to assist humans in geospatial data analysis tasks.
In individual cases, we observed a decline in performance after implementing GeoAgent. This drop is attributed to the additional noise introduced during the GeoAgent inference phases, which can be mitigated by incorporating an extra step in the process.
Among the three benchmarks, GeoCode-Others presents a greater challenge compared to the other two, as it has a zero completion rate for 7 out of 10 tasks in automatic mode. Despite this, GeoAgent significantly improved the pass rate for these tasks.
In contrast, CIBench is the easiest benchmark, achieving a 100\% completion rate for 6 out of 10 tasks.

\subsection{Discusssion}
In constructing our dataset, we considered various libraries for open-domain geospatial data analysis tasks. 
Choosing the right library is challenging and requires extensive expert knowledge. 
Our research shows that the GEE library is the only one that offers many resources within the open-source community, while other libraries mostly provide only usage examples. 
Including these libraries requires a lot of human effort. 
To ensure the fairness and validity of our experiments, we tested GEE separately from the other libraries. 
We generated code instructions from the provided code snippets for our dataset, although these instructions may not always match perfectly with codes. 
Additionally, prompt generation, which uses Llama3.1, highlights the importance of how prompts are worded to produce accurate, task-specific code. 
The representation of library and function dependencies is handled during the model's pre-training phase, but the exploration of these dependencies within prompts is limited. 
Further investigation into diverse ways to represent these dependencies in prompts could be an important area for future research.


In GeoAgent, using only Python interpreters for error detection is often not enough and can be misleading. 
In complex workflows, running code without errors does not guarantee its correctness. Our findings show that inaccurate execution feedback can limit the advantages of making dynamic adjustments. 
Currently, the dynamic adjustment strategies in MCTS focus too much on execution feedback. 
To address this problem, we need to create more flexible online evaluation strategies. Additionally, relying too heavily on execution feedback and RAG adds significant costs during model inference. 
Therefore, we must consider efficiency when integrating these methods into the MCTS framework. 
A possible solution is to implement selective execution feedback during dynamic adjustments instead of using it in every generation step.

In the evaluation of generated code, we identified four common types of errors. 
First, instruction-following errors occur when the model lacks adequate training in specific domain knowledge, leading to difficulties in following task instructions. 
Second, hallucination errors involve incorrect or non-existent function calls and undefined variables, often caused by using less common geospatial Python libraries. 
These errors reveal the challenges of applying LLMs to open-domain geospatial tasks. 
Third, the lack of information errors happens when the prompt does not provide enough or accurate information, which LLMs cannot fix on their own and need human help to address. Lastly, general code errors reflect the current limitations of LLMs in generating code effectively.

This work has two main limitations. 
First, GeoAgent relies solely on the capabilities of LLMs, while many open-source geospatial libraries could improve their geospatial coding skills. 
Second, the evaluation metrics for GeoCode focus only on process-oriented assessments. Although the code is executable, measuring its performance based on output is difficult because the results can be varied and complex.

\section{Conclusion}
We introduce GeoAgent, an innovative approach for geospatial task programming designed to enhance access to extensive geospatial datasets and facilitate automated workflow using LLMs with diverse and ever-evolving Python libraries. 
GeoAgent harnesses the capabilities of LLMs to transform tasks into executable units, subsequently retrieving the corresponding APIs through RAG from a knowledge database and dynamically refining subtasks in MCTS. 
Additionally, we develop a benchmark, GeoCode, to assess the efficacy of the proposed framework on diverse geospatial data analysis tasks using popular geospatial Python libraries. 
Our experimental results on GeoCode and existing benchmarks indicate that GeoAgent surpasses LLM baselines in both general programming tasks and geospatial data analysis tasks. 
The findings reveal that GeoAgent significantly enhances geospatial task pass rate and task completion.
With GeoAgent, we envision a future for advanced assistance tools that can seamlessly access relevant Python libraries and extensive online data for various geospatial tasks, thereby generating tailored code for researchers. 
We hope this work will contribute to improving the use of geospatial data in research aimed at societal benefits and environmental conservation.

\section*{Acknowledgements}
I would like to express my sincere gratitude to Professor Laurent Wendling for his valuable insights and guidance during the initial discussions of this work.

\bibliographystyle{IEEEtran}
\bibliography{sample-base.bib}

\newpage
\section{Appendices}
\subsection{Motivation}
This study aims to establish a benchmark for evaluating the capabilities of LLM coders in geospatial data analysis programming.
Current research has demonstrated the efficacy of natural language processing in remote sensing tasks.
Unlike LVLMs, geospatial data encompasses a multitude of modalities that are not readily accessible through natural language interfaces
Furthermore, the continuous updates in sensor technology and the corresponding Python libraries present challenges for users unfamiliar with these tools in data processing.
In this context, we seek a more foundational approach to processing diverse modalities and sources of geospatial data. 
Programming access to such data is a promising approach, as evidenced by advancements in NLP and CV communities.
The advantages of programming for geospatial data analysis include the availability of online data and the ability to execute code instantly and receive feedback. 
Moreover, most tasks can be completed using simple scripts without requiring cross-file programming.

Despite these advantages, there is a lack of comprehensive study of geospatial task programming using LLM agents. 
In this work, we focus particularly on the challenges associated with geospatial data analysis, namely: (1) the relatively low computational resources available for post-training, (2) the diversity of function calls, and (3) logical coherence in sequential tasks.
We also create a benchmark that can enhance our understanding of the potential of current LLM techniques in addressing these challenges and inform future research directions.
We constructed this dataset following three primary guidelines to encourage community contributions to this benchmark. 
First, each benchmark item should be executable, with feedback provided by the code interpreter instantaneously.
Benchmarks at the repository or single-function level are unsuitable for this purpose.
Second, the benchmark should be practical, encompassing a variety of geospatial data analysis scenarios, including image acquisition, spatial analysis, and beyond. 
Third, the evaluation should challenge the completion of tasks, requiring LLMs to demonstrate strong compositional reasoning and instruction-following capabilities.
Standalone API calls are inadequate for this purpose.
This initial effort aims to assist the community in better assessing the LLMs' programming capabilities in solving geospatial tasks within an open-ended framework, thereby contributing to the advancement of this research area.

\newpage
\subsection{Benchmark Construction Details}
\subsubsection{Python Library Version Control and Domain Classification}
\begin{longtable}{p{0.2\textwidth}p{0.3\textwidth}p{0.4\textwidth}}
\caption{
The version of Python libraries used in GeoCode, as well as their parent functionality domain.}
\label{tab9} \\ 
\hline
Library             & Version & Domain  \\
\hline
\endfirsthead

\hline
Library             & Version & Domain  \\
\hline
\endhead

\hline
\endfoot

\hline
earthengine-api     & 0.1.412 & RS data analysis \\
cubo                & 2024.8.0 & RS data acqusition \\
pystac              & 1.10.1 & RS data acquisition \\
eemont              & 0.3.6 & RS data analysis \\
geemap              & 0.34.0 & visualization \\
geetools            & 1.4.0 & RS data analysis \\
GemGIS              & 1.1.8 & GIS processing \\
Gempy               & 2024.2.0.2 & geological models \\
geeet               & 0.3.0 & evapotranspiration \\
GeoPandas           & 1.0.1 & GIS processing \\
GOES-2-Go           & 2024.7.0 & RS data acquisition  \\
GeoUtils            & 0.1.8 & raster and geographic information processing \\
gstools             & 1.6.0 & geostatistical \\
leafmap             & 0.37.1 & mapping and geospatial analysis \\
Lonboard            & 0.9.3 & visualizing large geospatial datasets  \\
meteostat           & 1.6.8 & weather and climate data  \\
pystac\_client      & 0.8.3 & RS data acquisition  \\
pytesmo             & 0.16.0 & Soil Moisture Observations  \\
scikeo              & 0.2.32 & machine learning\\
segment-geospatial  & 0.10.7 & deep learning\\
sen2nbar            & 2024.6.0  & Nadir BRDF Adjusted Reflectance\\
srai                & 0.7.6 & deep learning \\
planetary\_computer & 1.0.0 & RS data acquisition  \\
verde               & 1.8.1 & machine learning \\
wxee                & 0.4.2 & RS data analysis \\
xarray-spatial      & 0.4.0 & Raster-based Spatial Analytics \\
pylandtemp          & 0.0.1a1 & global land surface temperature\\
eradiate            & 0.28.0 & 3D radiative transfer model \\
spectramap          & 0.6.0.0 & Hyperspectral package for spectroscopists \\ \hline

\end{longtable}

\subsubsection{Code Collection:} 
All these datasets collected in this work are from publicly available GitHub repositories and solution-sharing websites. 
Initially, we collect a general corpus of geospatial-topic repositories for subsequent analysis.
This process begins by obtaining a list of geospatial data processing repositories through GitHub's search function, specifically targeting those with permissive licenses. 
Following the preliminary data collection, we extract usage examples from the Python projects. 
These examples, provided by the library authors, demonstrate the capabilities and use cases.
Among the geospatial data processing libraries, Google Earth Engine \cite{Gorelick2017GoogleEE} stands out as the most significant and widely used.
However, most existing scripts for this library are written in JavaScript. 
To harness these abundant code resources, we first employ LLMs to automatically translate JavaScript into Python scripts, followed by manual correction of any runtime errors through program analysis.
In addition to GitHub repositories, we also collect data from websites.
The acquisition of geospatial data is fundamental to effective geospatial data processing and analysis. 
Modern geospatial data platforms facilitate the collection of diverse datasets, which can be tailored to meet specific user requirements. 
As a result, we place a strong emphasis on integrating online data acquisition libraries into the proposed benchmark.

\subsubsection{Code Refactoring:} 
The majority of the code utilized in this benchmark originates from GEE, where it often includes commented code blocks and redundant and even dead codes. 
Consequently, the process of refactoring the collected code is crucial. 
The initial approach involves employing LLMs to automatically refactor the scripts, although LLMs frequently omit parts of the content when processing long contexts. 
To address this, the scripts are first divided into multiple steps based on code comments, after which each step is independently refactored into Python code.
Despite these efforts, the scripts still present several issues, such as unnecessary code blocks, unused imported libraries, and runtime errors.
To develop a high-quality execution-based benchmark, it is necessary to verify the correctness of each block, identify bugs, and remove redundant code and libraries.
However, this refactoring process is complex and time-consuming for human developers. 
To enhance code quality and reduce manual effort, a code interpreter and an abstract syntax tree parser are employed to analyze each code block.
The process involves anchoring each block, gathering all related code blocks to construct a script, and filtering out scripts with fewer than three executable units. 
At last, the code must pass the code interpreter test; otherwise, LLMs are re-engaged to refactor the code when it fails.
Additionally, unused libraries must be removed, and any necessary libraries that are missing in the code snippet should be added. 
When LLMs are involved in the refactoring loop, two issues are identified: first, the absence of essential variables in the code snippet, which prevents smooth execution, and second, the LLMs become stuck when resolving runtime bugs, leading to iterative refinement.
Continuous human feedback is crucial for providing viable solutions, and ensuring the process stays on track. 
The same methodology is applied to code from other Python libraries.

\subsubsection{Code Instruction Generation:} 
The evaluation of LLM coders relies on using the instruct-code pairs, where users pose questions or provide task-specific instructions to LLMs for task programming.
To construct instruct-code pairs, we initially process the code snippet through the Llama3.1 instruction version, which is prompted to produce an instruction for each code block. 
This approach results in step-by-step instructions that the LLM can then follow to generate a script using given libraries. 
Then, we manually ﬁx the instruction lacking the necessary information for corresponding code generation.
The process starts by iterating over each code block in the scripts.
The LLMs are then employed as summarizers to generate task descriptions for each block.
The summarization process is mathematically represented as follows:
\begin{equation} 
<s_i, S_i> = LLMs(c_i)
\end{equation}
where $s_i$ corresponds to the summarization of code snippet $c_i$ and
$S_i=\{s_1, ..., s_m\}$ refers to the description of previous blocks.
The summarization $s_i$  is considered as the subtask description of the code snippet $c_i$, while the aggregate description $S_i$ is treated as the overall task description.

\newpage
\subsection{Detailed Prompts}
\section*{}
\begin{promptbox}[Prompt for Program Refactoring]
Here is a task script written in JavaScript using GEE; please refactor it into a Python version using the Earth Engine API.
\end{promptbox}

\vspace{0.1 cm} 

\begin{promptbox}[Prompt for instruction generation]
Here are the used Python libraries \{ \} and the previous steps \{ \}. I want you to become my Expert Prompt Creator. Your goal is to help me reverse Python code into prompts. The prompt you provide should be a very detailed text description, including all necessary information, such as exact parameter values, input files, date, location, etc. Provide the prompt for this step's code \{ \} in the format \{ 'prompt': detailed prompts with all parameters and values \}.
\end{promptbox}

\vspace{0.1 cm} 

\begin{promptbox}[Instruction-tuning Prompt]
I will give you a task description that needs to use some of the given library \{ \} to implement it. You need to provide the Python code for the given task description. Here is the previous code \{ \}. Please provide the Python code for the current step with this description.
\end{promptbox}

\vspace{0.1 cm} 

\begin{promptbox}[Inference Prompt]
I want you to become my expert programmer. Your goal is to help me write Python code for the given task using the Python library \{ \}. You need to write code according to the detailed prompt \{ \}. Please provide the corresponding code.
\end{promptbox}

\vspace{0.1 cm} 

\begin{promptbox}[Updating the Node Prompt]
Your goal is to fix the error in the initial prompt. I first give you the Python code \{ \}. Give your solution for fixing the error \{ \} and add it to the initial prompt.
\end{promptbox}

\vspace{0.1 cm} 

\begin{promptbox}[Prepend New Task Prompt]
Defining the undefined variables for the next step task: \{ next step instruction \}. Give your code for the undefined variables in this step:
\end{promptbox}

\newpage
\subsection{Benchmark Samples}

\section*{}
\begin{promptbox}[Traning examples]
\begin{lstlisting}[style=python, breaklines=true,breakatwhitespace=true]
(*@\#\# \textbf{Instruction}@*)
# Collect Sentinel-2 data for the period from June 1st, 2019 to September 30th, 2019, and filter out images with more than 20% cloudy pixels. Then, collect Sentinel-2 data for the period from June 1st, 2021 to September 30th, 2021, and filter out images with more than 20% cloudy pixels. Use the "maskS2clouds" function to mask out cloudy pixels in both data collections.
(*@\#\# \textbf{Ground Truth}@*)
import ee
# Sentinel-2 data collection
Date_1 = ee.ImageCollection('COPERNICUS/S2_SR_HARMONIZED') \
                  .filterDate('2019-06-01', '2019-09-30') \
                  .filter(ee.Filter.lte('CLOUDY_PIXEL_PERCENTAGE',20)) \
                  .map(maskS2clouds)
Date_2 = ee.ImageCollection('COPERNICUS/S2_SR_HARMONIZED') \
                  .filterDate('2021-06-01', '2021-09-30') \
                  .filter(ee.Filter.lte('CLOUDY_PIXEL_PERCENTAGE',20)) \
                  .map(maskS2clouds)
\end{lstlisting}
\end{promptbox}

\begin{promptbox}[Evaluation Cases]
\begin{lstlisting}[style=python, breaklines=true,breakatwhitespace=true]
(*@\#\# \textbf{Instruction}@*)
#1. Calculate the median value of Date_1 for S1_img; 2. Clip S1_img using the ROI defined by "seoul"; 3. Calculate the median value of Date_2 for S2_img; 4. Clip S2_img using the ROI defined by "seoul"; 5. Combine the clipped S1_img and S2_img to create a composite image. Here is the previous code:
import ee
# Sentinel-2 data collection
Date_1 = ee.ImageCollection('COPERNICUS/S2_SR_HARMONIZED') \
                  .filterDate('2019-06-01', '2019-09-30') \
                  .filter(ee.Filter.lte('CLOUDY_PIXEL_PERCENTAGE',20)) \
                  .map(maskS2clouds)

Date_2 = ee.ImageCollection('COPERNICUS/S2_SR_HARMONIZED') \
                  .filterDate('2021-06-01', '2021-09-30') \
                  .filter(ee.Filter.lte('CLOUDY_PIXEL_PERCENTAGE',20)) \
                  .map(maskS2clouds)

\end{lstlisting}
\end{promptbox}

\subsection{Error Samples}
\section*{}
\begin{promptbox}[Instruction Following Errors]
\begin{lstlisting}[style=python, breaklines=true,breakatwhitespace=true]
(*@\#\# \textbf{Instruction}@*)
# Import Necessary Feature Collections and Image Collections:
# Import the Lower Deschutes feature collection.
# Set visualization parameters for the annual NPP data: [Opacity: 1, Bands: ["annualNPP"], Min: 2178.6000000000004, Max: 10727.4, Gamma: 1]
# Import the Landsat 30m NPP image collection.
# Import the MODIS 250m NPP image collection.

(*@\#\# \textbf{Generation}@*)
# Import necessary feature collections and image collections
Lower_Deschtues = ee.FeatureCollection("users/yusin/Lower_Deschtues")
# Set visualization parameters for the annual NPP data
Map.addLayer(Annual_NPP, {'opacity':1, 'bands':["annualNPP"],'min':2178.6000000000004,'max':10727.4, 'gamma':1}, "Annual NPP")
Landsat_30m_NPP = ee.ImageCollection("LANDSAT/LC08/C01/T1_8DAY_NPP")
MODIS_250m_NPP = ee.ImageCollection("MODIS/006/MOD17A2HGF")

(*@\#\# \textbf{Error Feedback}@*)
(*@{\color{YC}$NameError:~name~'Annual\_NPP'~is~not~defined$}@*)

(*@\#\# \textbf{Ground Truth}@*)
# Import the Lower Deschutes feature collection
lower_deschutes_fc = ee.FeatureCollection('users/your_username/Lower_Deschutes')
# Set visualization parameters for the annual NPP data
npp_vis_params = {
  'opacity': 1,
  'bands': ['annualNPP'],
 'min': 2178.6,
 'max': 10727.4,
  'gamma': 1
}
# Import the Landsat 30m NPP image collection
landsat_npp_collection = ee.ImageCollection('LANDSAT/LT05/C01/T1_TOA')
# Import the MODIS 250m NPP image collection
modis_npp_collection = ee.ImageCollection('MOD11C3')
\end{lstlisting}
\end{promptbox}

\vspace{0.1 cm} 

\begin{promptbox}[Non-existent APIs]
\begin{lstlisting}[style=python, breaklines=true,breakatwhitespace=true]
(*@\#\# \textbf{Instruction}@*)
# Clip the buildings dataset to the geometry of the selected region.

(*@\#\# \textbf{Generation}@*)
#Clipping the Buildings Dataset:
clipped_buildings = buildings.clip(geometry)
Map.addLayer(clipped_buildings, {}, 'Google Open Buildings')

(*@\#\# \textbf{Error Feedback}@*)
(*@{\color{YC}AttributeError: 'FeatureCollection' object has no attribute 'clip'}@*)

(*@\#\# \textbf{Ground Truth}@*)
# Clip the buildings dataset to the region of interest
buildings_clipped = buildings.map(lambda img: img.clip(roi))
\end{lstlisting}
\end{promptbox}

\vspace{0.1 cm} 

\begin{promptbox}[Hallucination of API parameters]
\begin{lstlisting}[style=python, breaklines=true,breakatwhitespace=true]
(*@\#\# \textbf{Instruction}@*)
# Utilize ee.Algorithms.Landsat.simpleComposite to create a simple composite image:- Collection: `l8raw` - Parameter: `asFloat = true`

(*@\#\# \textbf{Generation}@*)
image = ee.Algorithms.Landsat.simpleComposite({
  'collection': l8raw,
  'asFloat': True
})
Map.addLayer(image)

(*@\#\# \textbf{Error Feedback}@*)
(*@{\color{YC}EEException: Unrecognized argument type to convert to an ImageCollection: {'collection': <ee.imagecollection.ImageCollection object at 0x7f7e384acac0>, 'asFloat': True}@*)

(*@\#\# \textbf{Ground Truth}@*)
# Utilize ee.Algorithms.Landsat.simpleComposite to create a simple composite image
composite_image = ee.Algorithms.Landsat.simpleComposite(l8raw, asFloat=True)
\end{lstlisting}
\end{promptbox}

\vspace{0.1 cm} 

\begin{promptbox}[Undefined Variables]
\begin{lstlisting}[style=python, breaklines=true,breakatwhitespace=true]
(*@\#\# \textbf{Instruction}@*)
# Filter and Aggregate Global Fishing Effort by Date Range: - Start Date: `2016-01-01` - End Date: `2017-01-01` - The dataset is filtered by the date range and the global flag state ('WLD'). - The filtered collection is aggregated to a single image representing the total fishing effort.

(*@\#\# \textbf{Generation}@*)
gfwFiltered = gfw \
 .filterDate(startDate, endDate) \
 .filter(ee.Filter.eq('flag', 'WLD'))
gfwFilteredTotal = gfwFiltered.sum()

(*@\#\# \textbf{Error Feedback}@*)
(*@{\color{YC}NameError: name 'startDate' is not defined}@*)

(*@\#\# \textbf{Ground Truth}@*)
start_date = '2016-01-01'
end_date = '2017-01-01'
gfwFiltered = gfw \
.filterDate(start_date, end_date)\
.filter(ee.Filter.eq('flag_state', 'WLD'))
gfwFilteredTotal = gfwFiltered.reduce(ee.Reducer.sum())
\end{lstlisting}
\end{promptbox}

\begin{promptbox}[Syntax Errors]
\begin{lstlisting}[style=python, breaklines=true,breakatwhitespace=true]
(*@\#\# \textbf{Instruction}@*)
# A polygon defined by the coordinates [[11.3258, 47.0801], [11.6746, 47.1773], [11.5318, 47.3730], [11.1692, 47.2687]].

(*@\#\# \textbf{Generation}@*)
geometry = ee.Geometry.Polygon(
        [[11.3258, 47.0801], [11.6746, 47.1773], 
        [11.5318, 47.3730],[11.1692, 47.2687]]])

(*@\#\# \textbf{Error Feedback}@*)
(*@{\color{YC}SyntaxError: closing parenthesis ']' does not match opening parenthesis '('}@*)

(*@\#\# \textbf{Ground Truth}@*)
geometry = ee.Geometry.Polygon(
    [[11.3258, 47.0801], [11.6746, 47.1773],
     [11.5318, 47.3730],[11.1692, 47.2687]])
\end{lstlisting}
\end{promptbox}

\newpage
\subsection{Example Studies}
Our benchmark introduces instances that facilitate geospatial data analysis. 
While the practical applications of this work have not yet been fully explored, our audience may be eager to understand its potential impact on the geoscience community.
In this section, we highlight a change detection use case utilizing Sentinel-1 GRD images in conjunction with a log-ratio thresholding algorithm. 
The presented example involves instructions, failed generation, error feedback, passed generation, successful generation, self-fixed generation, and human intervention (when it goes to be unsolved) in terms of each task programming.
This demonstration not only showcases the current capabilities and limitations of GeoAgent but also provides an evaluative perspective for the audience.

\section*{}
\begin{promptbox}[Change Detection]
\begin{lstlisting}[style=python, breaklines=true,breakatwhitespace=true]
(*@\#\# \textbf{Instruction 1:}@*)
# Import library GEE python and geemap, then authenticate with GEE; at last start a Map instance.
(*@\#\# \textbf{Sucessful Generation:}@*)
import ee
ee.Authenticate()
import geemap
Map = geemap.Map()
(*@\#\# \textbf{Instruction 2:}@*) 
# Define Parameters:\n\nDate Range: - Start Date: '2018-01-01' - End Date: '2019-01-01'\nArea of Interest (AOI): Polygon with coordinates defining the area [[-122.0, 37.0],[-122.0, 38.0],[-121.0, 38.0],[-121.0, 37.0],[-122.0, 37.0]].
(*@\#\# \textbf{Sucessful Generation:}@*)
start_date = '2018-01-01'
end_date = '2019-01-01'
aoi = ee.Geometry.Polygon([[-122.0, 37.0], [-122.0, 38.0], [-121.0, 38.0], [-121.0, 37.0], [-122.0, 37.0]])
(*@{\color{YC}$Map.addLayer(ee.ImageCollection(MODIS/006/MOD13Q1).filterDate(start\_date, end\_date)$}@*)
(*@{\color{YC}$.filterBounds(aoi_polygon).first())$}@*)
(*@\#\# \textbf{Instruction 3:}@*) 
#Select Sentinel-1 Image Collection:'COPERNICUS/S1_GRD'; Filter Criteria: Date Range: '2018-01-01' to '2019-01-01', Orbit Properties: 'DESCENDING', Geographical Bounds: Defined by aoi; Sort sentinel-1 image according to system. 
(*@\#\# \textbf{Sucessful Generation:}@*)
sentinel1_data = ee.ImageCollection('COPERNICUS/S1_GRD').filterDate(start_date, end_date).filter(ee.Filter.eq('orbitProperties_pass', 'DESCENDING'))        .filterBounds(aoi)
collection_S1data = sentinel1_data.sort('system:time_start')
(*@\#\# \textbf{Instruction 4:}@*)
#Define box kernel Kernel Size: 33 meters Write a Python function named smooth that takes an input image. The function should: Revert the input image to its original form by dividing it by 10 and raising it to the power of 10. Apply a boxcar convolution filter to the reverted image to smooth it. Convert the filtered image back to the backscatter coefficient by taking the base-10 logarithm and multiplying it by 10. Ensure that the resulting image retains the system:time_start and cycleNumber properties from the input image. Return the final image as a float type.
(*@\#\# \textbf{Failed Generation:}@*)
def smooth(image):
    reverted_image = ee.Image(image).divide(10).pow(10)
    (*@{\color{YC}$box\_kernel~=~ee.Kernel.box(33)$}@*)
    smoothed_image = reverted_image.focal(box_kernel, 'gauss', 1)
    log_image = smoothed_image.log10().multiply(10)
    filtered_image = log_image.addBands(image.select([0, 1, 2, 3]).copyProperties(image, ['system:time_start', 'cycleNumber']))
    return filtered_image
smoothed_collection = collection_S1data.map(smooth)
(*@\#\# \textbf{Error Feedback}@*)
(*@{\color{YC}AttributeError: type object 'Kernel' has no attribute 'box'}@*)
(*@\#\# \textbf{Human Intervention:}@*)
boxcar = ee.Kernel.square(33, 'meters')
def smooth(image):
    natimg = ee.Image(10.0).pow(image.divide(10.0))
    filtered = natimg.convolve(boxcar)   
    return filtered.log10().multiply(10.0).set('system:time_start', image.get('system:time_start')).set('cycleNumber', image.get('cycleNumber')).float()
collection_S1data_Filtered = collection_S1data.map(smooth)
(*@\#\# \textbf{Instruction 5:}@*)
#Create a visualization using the Google Earth Engine (GEE) Map object to display Sentinel-1 Ground Range Detected (GRD) data before and after a filtering step. The code should add two layers to the map: The first layer should represent the unfiltered Sentinel-1 data. Use the first image from the collection_S1data ImageCollection. The visualization parameters should specify the VH band with a minimum value of -23 and a maximum value of -11. Label this layer as Sentinel-1 GRD non filtered. The second layer should represent the filtered Sentinel-1 data. Use the first image from the collection_S1data_Filtered ImageCollection. The visualization parameters should also specify the VH band with a minimum value of -23 and a maximum value of -11. Label this layer as Sentinel-1 GRD filtered.The layers should be added to the map in the order specified.
(*@\#\# \textbf{Sucessful Generation:}@*)
Map.addLayer(collection_S1data.first().select('VH'), {'min': -23, 'max': -11}, 'Sentinel-1 GRD non filtered')
Map.addLayer(collection_S1data_Filtered.first().select('VH'), {'min': -23, 'max': -11}, 'Sentinel-1 GRD filtered')
(*@\#\# \textbf{Instruction 6:}@*)
#Write a Python function that adds a new band to an image, representing the ratio between the 'VH' and 'VV' bands from Sentinel-1 satellite data. The function should compute the ratio as the difference between the 'VH' and 'VV' bands, convert it to a floating-point number, and then add it as a new band named 'ratio'. After defining this function, apply it to a pre-existing image collection named collection_S1data_Filtered, mapping it across the entire collection to create a new image collection named collection_S1data_Filtered_Ratio.
(*@\#\# \textbf{Passed Generation:}@*)
def add_ratio_band(image):
    ratio = image.select('VH').subtract(image.select('VV'))\
    .divide(image.select('VV')).multiply(100)
    ratio_image = image.addBands(ratio.rename('ratio'))
    return ratio_image
collection_S1data_Filtered_Ratio = collection_S1data_Filtered.map(add_ratio_band)
(*@\#\# \textbf{Human Intervention}@*)
def add_vhvv_ratio(image):
    vh_vv_ratio = image.select('VH').subtract(image.select('VV')).float().rename('ratio')
    image = image.addBands(vh_vv_ratio) 
    return image
collection_S1data_Filtered_Ratio = collection_S1data_Filtered.map(add_vhvv_ratio)
(*@\#\# \textbf{Instruction 7:}@*)
#Generate Python code using the Google Earth Engine (GEE) Python API to visualize Sentinel-1 data. Convert the first 30 images in collection_S1data_Filtered_Ratio to a list using the toList() method. Assign this list to the variable S1_listData. Visualize the first image in this list on a map, clipped to the Area of Interest (aoi). Display the image in RGB format using the VV, VH, and ratio bands. Set the visualization parameters with minimum values of [-18, -23, 3] and maximum values of [-4, -11, 15]. Add this visualization to the map with the label sentinel-1 GRD RGB First. Repeat the visualization for the last image in the list, adding it to the map with the label Sentinel-1 GRD RGB Last.
(*@\#\# \textbf{Failed Generation:}@*)
first_image = S1_listData.get(0)
last_image = S1_listData.get(-1)
first_image_clipped = first_image.clip(aoi)
Map.addLayer(first_image_clipped.select('VV', 'VH', 'ratio'), {'min': [-18, -23, 3], 'max': [-4, -11, 15]}, 'sentinel-1 GRD RGB First')
last_image_clipped = last_image.clip(aoi)
S1_listData = collection_S1data_Filtered_Ratio.limit(30).toList(30)
Map.addLayer(first_image.clip(aoi), {'min': [-18, -23, 3], 'max': [-4, -11, 15]}, 'sentinel-1 GRD RGB First')
Map.addLayer(last_image.clip(aoi), {'min': [-18, -23, 3], 'max': [-4, -11, 15]}, 'Sentinel-1 GRD RGB Last')
(*@\#\# \textbf{Error Feedback}@*)
(*@{\color{YC}$NameError:~name~'S1\_listData'~is~not~defined$}@*)
(*@\#\# \textbf{Self-fixed Generation:}@*)
S1_listData = collection_S1data_Filtered_Ratio.toList(30)
first_image = ee.Image(S1_listData.get(0))
last_image = ee.Image(S1_listData.get(-1))
first_image_clipped = first_image.clip(aoi)
Map.addLayer(first_image_clipped.select('VV', 'VH', 'ratio'), {'min': [-18, -23, 3], 'max': [-4, -11, 15]}, 'sentinel-1 GRD RGB First')
last_image_clipped = last_image.clip(aoi)
Map.addLayer(ee.Image(S1_listData.get(29)).clip(aoi),{'bands': ['VV', 'VH', 'ratio'],min: [-18, -23, 3], max: [-4, -11, 15]}, 'Sentinel-1 GRD RGB Last')
(*@\#\# \textbf{Instruction 8:}@*)
#The first step is to select the Sentinel-1 radar images before and after the event from the provided list (S1_listData). Both images are clipped to a specified area of interest (aoi). Specifically, select the VH polarization band from the images. S1_before is the image before the event, and it's selected as the first image in the list. S1_after is the image after the event, selected as the 30th image in the list. Calculation of Log Ratio: Compute the Log Ratio by subtracting the S1_after image from the S1_before image. The result is then divided by 10 to normalize the values, making the output more scalable. This operation results in a new band named 'logratio'. Visualization: Finally, visualize the resulting Log Ratio image on the map. Set the visualization parameters with a min value of -1.5 and a max value of 1.5.
(*@\#\# \textbf{Sucessful Generation:}@*)
S1_before = ee.Image(S1_listData.get(0)).select('VH').clip(aoi)
S1_after = ee.Image(S1_listData.get(-1)).select('VH').clip(aoi)
LogRatio = S1_before.subtract(S1_after).divide(10).rename('logratio')
print('Log Ratio Band (Change Image):', LogRatio)
Map.addLayer(LogRatio, {'bands': ['logratio'], min: [-1.5], max: [1.5]}, 'Sentinel-1 Log-Ratio Image')
(*@\#\# \textbf{Instruction 9:}@*)
#Extract the first image from the Sentinel-1 dataset (collection_S1data). Center the map on the area of interest (aoi) with a zoom level of 10. Add the original Sentinel-1 image layer to the map with a visualization range (min: -20, max: 0). 
#Display Filtered Data: Filter the smoothed Sentinel-1 dataset (collection_S1data_Filtered) to retrieve images within a specified date range (start_date to end_date). Extract the first image from the filtered dataset. Add this filtered image layer to the map with the same visualization range (min: -20, max: 0). 
#Display VH/VV Ratio Data (collection_S1data_Filtered_Ratio): Extract the VH/VV ratio band from the first image of the filtered smoothed dataset. Extract the VH/VV ratio band from the last image of the filtered smoothed dataset, sorted by the system:index.\nAdd these VH/VV ratio layers to the map with a visualization range (min: 0, max: 1). Label the layers as VH/VV Ratio (First Image) and VH/VV Ratio (Last Image) respectively. 
(*@\#\# \textbf{Failed Generation:}@*)
filtered_S1_layer = collection_S1data.select('VH').clip(aoi)
Map.addLayer(filtered_S1_layer, {'bands': ['VH'], 'min': [-20], 'max': [0]}, 'Filtered Sentinel-1 Image')
vh_vv_ratio_first_image = ee.Image(collection_S1data_Filtered.get(0)).select('vhvv_ratio')
vh_vv_ratio_last_image = ee.ImageCollection(collection_S1data_Filtered).filterDate(start_date, end_date).sort('system:index').last().select('vhvv_ratio')
Map.addLayer(vh_vv_ratio_first_image.clip(aoi), {'bands': ['vhvv_ratio'], 'min': [0], 'max': [1]}, 'VH/VV Ratio (First Image)')
Map.addLayer(vh_vv_ratio_last_image.clip(aoi), {'bands': ['vhvv_ratio'], 'min': [0], 'max': [1]}, 'VH/VV Ratio (Last Image)')
(*@\#\# \textbf{Error Feedback}@*)
(*@{\color{YC}$AttributeError:~'ImageCollection'~object~has~no~attribute~'clip'$}@*)
(*@\#\# \textbf{Human Intervention}@*)
original_image = collection_S1data.first()
Map.setCenter(aoi.getCenter(), 10)
Map.addLayer(original_image, {'min': -20,'max': 0}, 'Original Data')
filtered_image =collection_S1data_Filtered.filterDate(start_date, end_date).first()
Map.addLayer(filtered_image, {'min': -20,'max': 0}, 'Filtered Data')
first_image = collection_S1data_Filtered_Ratio.filterDate(start_date, end_date).first()
last_image = collection_S1data_Filtered_Ratio.filterDate(start_date, end_date).sort('system:index').last()
Map.addLayer(first_image.select('VH_VV_ratio'), {'min': 0,'max': 1}, 'VH/VV Ratio (First Image)')
Map.addLayer(last_image.select('VH_VV_ratio'), {'min': 0,'max': 1}, 'VH/VV Ratio (Last Image)')
(*@\#\# \textbf{Instruction 10:}@*)
#Create a binary image by applying a threshold on LogRatio, where values greater than 1 are classified as change (True) and others as no change (False). Add the binary change detection layer to the map using the Map.addLayer() function. The visualization should use the 'logratio' band with values ranging from 0 to 1.
(*@\#\# \textbf{Passed Generation:}@*)
vh_vv_ratio_image_logratio = LogRatio.gt(1).log().rename('logratio')
Map.addLayer(vh_vv_ratio_image_logratio, {'min': 0, 'max': 1}, 'VH/VV Ratio (Logratio)')
(*@\#\# \textbf{Human Intervention}@*)
SARchange = LogRatio.gt(1)
Map.addLayer(SARchange, {'bands': ['logratio'], min: [0], max: [1]}, 'Sentinel-1 Log-Ratio Image')

\end{lstlisting}
\end{promptbox}

\newpage
\subsection{Ethics Statement}
Our benchmark is developed using Llama3.1 and a variety of online resources. 
It is carefully propossed by experts to ensure the exclusion of any private information.
Furthermore, we utilize GPT-4o for refining text and correcting typographical errors and syntax in both the benchmark and paper writing. 
Although Llama3.1 is associated with ethical risks, such as the potential generation of harmful content, all generated instructions within this benchmark undergo human review, and the code is executed in sandbox environments to mitigate these risks. 
LLMs are known for their high computational demands, which pose environmental concerns. 
However, our primary work does not involve training such a large model; instead, it involves only inference on benchmarks, which results in a relatively low computational footprint. 
All experiments conducted for this study were performed on a single NVIDIA RTX 4090 machine.
\end{document}